# Machine-learning guided discovery of a high-performance spin-driven thermoelectric material

Yuma Iwasaki[1,2]*, Ichiro Takeuchi[3,4], Valentin Stanev[3,4], Aaron Gilad Kusne[3,5], Masahiko Ishida[1], Akihiro Kirihara[1], Kazuki Ihara[1], Ryohto Sawada[1], Koichi Terashima[1], Hiroko Someya[1], Ken-ichi Uchida[2,6,7,8,9], Shinichi Yorozu[1] and Eiji Saitoh[8,9,10,11]

[1]Central Research Laboratories, NEC Corporation, Tsukuba 305-8501, Japan

[2]PRESTO, JST, Saitama 322-0012, Japan

[3]Department of Materials Science and Engineering, University of Maryland, College Park, MD 20742, USA

[4]Center for Nanophysics and Advanced Materials, University of Maryland, College Park, MD 20742, USA

[5]National Institute of Standards and Technology, Gaithersburg, MD 20899, USA

[6]Research Center for Magnetic and Spintronic Materials (CMSM), National Institute for Materials Science (NIMS), Tsukuba, 305-0047, Japan

[7]Research and Services Division of Materials Data and Integrated System (MaDIS), National Institute for Materials Science (NIMS), Tsukuba, 305-0047, Japan

[8]Institute for Materials Research, Tohoku University, Sendai 908-8577, Japan

[9]Center for Spintronics Research Network, Tohoku University, Sendai, 980-8577, Japan

[10]Advanced Institute for Materials Research, Tohoku University, Sendai 908-8577, Japan

[11]Advanced Science Research Center, Japan Atomic Energy Agency, Tokai, 319-1195, Japan

Email: [y-iwasaki@ih.jp.nec.com](mailto:y-iwasaki@ih.jp.nec.com)

**Thermoelectric conversion using Seebeck effect for generation of electricity is becoming an indispensable technology for energy harvesting and smart thermal management. Recently, the spin-driven thermoelectric effects (STEs), which employ emerging phenomena such as the spin-Seebeck effect (SSE) and the anomalous Nernst effect (ANE), have garnered much attention as a promising path towards low cost and versatile thermoelectric technology with easily scalable manufacturing. However, progress in development of STE devices is hindered by the lack of understanding of the mechanism and materials parameters that govern the STEs. To address this problem, we enlist machine learning modeling to establish the key physical parameters controlling SSE. Guided by these models, we have carried out a high-throughput experiment which led to the identification of a novel STE material with a thermopower an order of magnitude larger than that of the current generation STE devices.**

Waste heat is ubiquitous in modern society, and thermoelectric technologies based on the Seebeck effect have been embraced as a key avenue to a sustainable future[1-3]. Unfortunately, conventional thermoelectric (TE) devices suffer from high fabrication cost due to their complex structure, in which p-type and n-type thermoelectric materials are cascade-connected in an alternating way. The emergence of novel thermoelectric devices based on the spin-driven thermoelectric (STE) phenomena offers a potential solution to this problem. In contrast to the conventional TE devices, the STE devices consist of simple layered structures, and can be manufactured with straightforward processes, such as sputtering, coating and plating, resulting in lower fabrication costs[4]. An added advantage of the STE devices is that they can double their function as heat-flow sensors, owing to their flexible structures and lower thermal resistance [5].

STE devices utilize the spin-Seebeck effect (SSE) [6-10] and the anomalous Nernst effect (ANE)[11-13]. SSE generates a spin current from a temperature gradient in a magnetic material. By connecting a metallic film having large spin-orbit interaction (such as Pt) to a magnetic material, one can convert the spin current into the electrical current via the inverse spin-Hall effect (ISHE) [14-17]. Thermoelectric conversion based on SSE can lead to an entirely new class of inexpensive and versatile thermoelectric devices [4]. Unfortunately, advance in device development is hampered by the lack of understanding of the fundamental mechanism behind SSE and the materials parameters governing it. Several different theories have been put forth to explain the phenomenon[18,19], but a unified picture of its mechanism is yet to emerge. Key materials parameters driving SSE have not been identified to date, and there are no clear pathways to enhance the thermopower and related figures of merit.

To address this problem, we have developed a systematic approach to uncovering the major materials variables governing the SSE. We combine machine learning modeling with high-throughput experimentation, and use modeling results for designing combinatorial libraries[20-24]. We have successfully leveraged the machine-learning-informed knowledge of the dependence of SSE on materials parameters to arrive at a novel and high-performance STE material utilizing ANE, which converts a heat current into an electrical current via the spin-orbit interaction in a single ferromagnetic material. Out of a number of proposed materials systems, a composition spread of one ternary system has led to the identification of $Fe_{0.665}Pt_{0.27}Sm_{0.065}$, which exhibits thermopower as large as 11.12 $\mu$V/K.

**Results**

**Primary experiment.** Schematic of the SSE device used in the primary experiment is shown in Figure 1a. It is composed of a paramagnetic conductive layer, a magnetic layer, and a single crystal substrate. We adopted a bilayer consisting of platinum (Pt) and rare-earth-substituted yttrium iron garnet ($R_1Y_2Fe_5O_{12}$, (R:YIG)), where R stands for a rare-earth element. The YIG and Pt are believed to be a good combination for STE conversion due to their long magnon-diffusion length ($D_L$) and large spin-orbit interaction[4]. Initially, we focused our attention on the effect of different rare-earth elements (in YIG) on SSE.

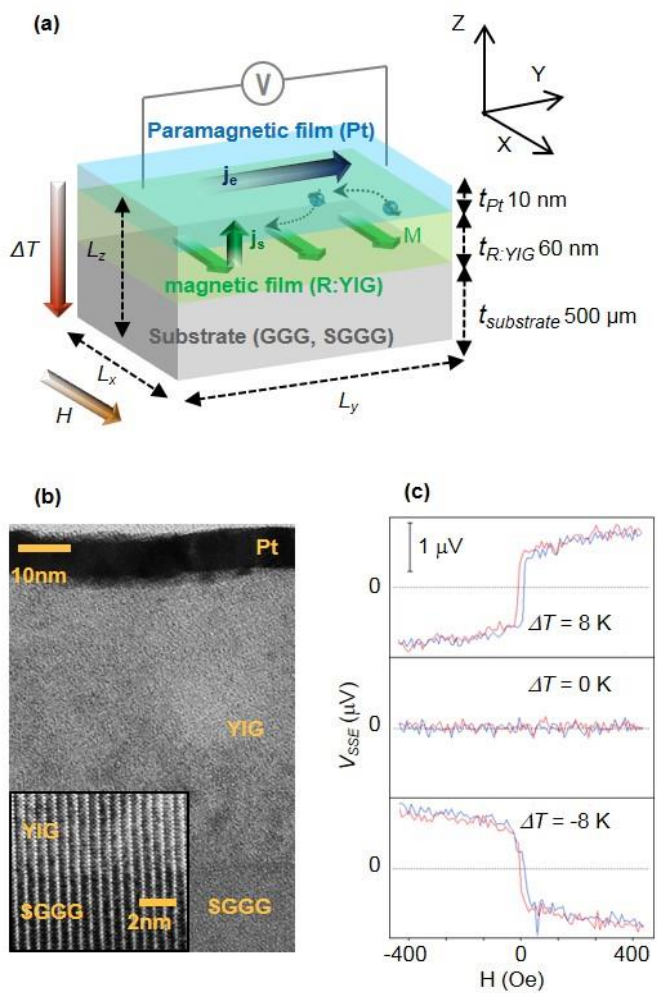

**Figure 1**

The typical thickness of the R:YIG layer is 60 nm. Since strain in the films is known to influence spin transport in materials and because the lattice constant of R:YIG depends on the choice of R, two substrates were used: a (111)-oriented Gadolinium Gallium Garnet ($Gd_3Ga_5O_{12}$, referred to as GGG, with the lattice constant $a$ = 12.385 Å) and a (111)-oriented Substituted Gadolinium Gallium Garnet ($Gd_{2.675}Ca_{0.325}Ga_{4.025}Mg_{0.325}Zr_{0.65}O_{12}$, referred to as SGGG, with the lattice constant $a$ = 12.464 Å). Thus, different R:YIG layers have different degrees of lattice match to the two substrates. Figure 1b shows the cross-sectional image of a typical device observed by transmission electron microscopy. The epitaxial YIG layer with a coherent interface was formed on the SGGG substrate. When a temperature difference $\Delta T$ and a magnetic field $H$ are applied along the z and x direction, respectively, a spin current is generated along the z direction by the spin-Seebeck effect in the R:YIG layer (Figure. 1a). The spin current is then injected into the Pt layer and converted into an electric current by the ISHE. One then detects the thermoelectric voltage (SSE voltage, $V_{SSE}$) along the y direction. Figure 1c shows typical $V_{SSE}$ behaviors as a function of $H$ for different $\Delta T$. Note that the sign of $V_{SSE}$ changes when the sign of $H$ is inverted - a clear indication that the thermoelectric voltage arises from the SSE and the ISHE.

The substitution of the c-site in R:YIG with different rare-earth elements changes the physical properties of the material, and allows us to study the impact this has on the STE phenomena. Figure 2 shows the thermopower [25] $S_{SSE}(\equiv (V_{SSE}/\Delta T)(L_z/L_y))$ measured for Pt/R:YIG samples, fabricated on GGG or SGGG substrates and with rare-earth elements R - La, Ce, Pr, Nd, Sm, Eu, Gd, Tb, Dy, Ho, Er, Tm, Yb, and Lu (Pm, which is radioactive, was excluded from the study). It is clear that $S_{SSE}$ strongly depends on the choice of the R element. Differences in measured $S_{SSE}$ can be dramatic; for example, the response of Pt/Yb:YIG/GGG is about three times as large as that of Pt/YIG/GGG.

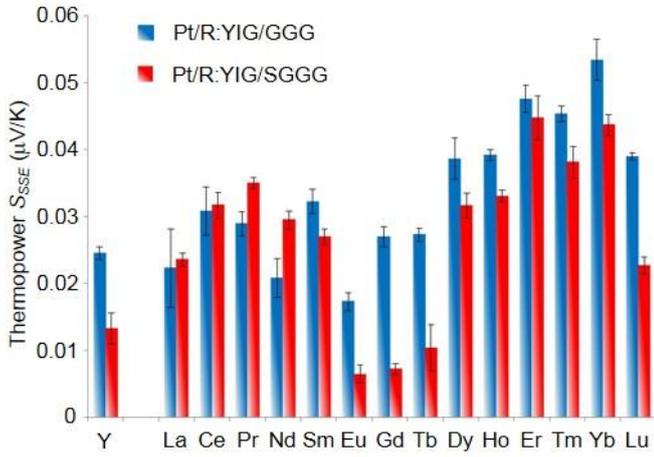

**Figure 2**

There is also a more subtle dependence of $S_{SSE}$ on the choice of substrate. This must arise from a secondary factor, since GGG and SGGG substrates are non-magnetic and thus cannot directly impact SSE. We believe that $S_{SSE}$ is influenced by strain and the crystalline quality of R:YIG, which depend on the lattice mismatch between R:YIG and the substrate. Indeed, Figure. 2 indicates that R:YIG layers with large ionic radius of the R element (light rare earth element) on SGGG tend to generate larger $S_{SSE}$ than the same R:YIG layers on GGG. We attribute this to the smaller lattice mismatch between R:YIG and SGGG (SGGG has a larger lattice constant than GGG). For instance, Pt/La:YIG/GGG generates smaller $S_{SSE}$ than Pt/La:YIG/SGGG. The pattern is reversed for the heavier rare earth elements; Pt/Lu:YIG on GGG generates larger $S_{SSE}$ than on SGGG.

The striking dependence of $S_{SSE}$ on the choice of R shown in Figure 2 suggests that the physical parameters of R strongly influence the SSE. To analyze and quantify this dependence, we need descriptors that can encode the properties of different rare-earth elements, such as atomic weight $n_R$, spin and orbital angular momenta $S_R$ and $L_R$, number of unfilled orbitals, elemental melting temperatures, magnetic moments, and ground state volumes, etc. We also consider the lattice mismatch $\Delta a$ between R:YIG and the substrate. It is, however, difficult to experimentally isolate and extract the $S_{SSE}$ dependence on a given physical parameter of R. To delineate the relation between the atomic weight $n_R$ and $S_{SSE}$, for instance, it would be necessary to measure the $S_{SSE}$ for different $n_R$ while keeping all other predictors fixed, which is experimentally not feasible. In order to uncover the subtle correlations and the physical origin of the SSE hidden in the initial experimental results, we used supervised machine learning[26].

**Machine learning modeling.** We employ three types of supervised machine learning models: Elastic Net (EN), Quadratic Polynomial LASSO (QP-LASSO), and Neural Network (NN) [26]. The EN model is constrained to only linear dependence, but it is straightforward to apply for extracting dominant descriptors. In contrast, the NN model is very flexible and can reproduce a highly non-linear dependence, but is difficult to interpret and understand. The QP-LASSO is in between NN and EN in terms of complexity and interpretability. In order to reduce the risk of over-fitting the available experimental data, we fix the number of descriptors to four, namely $\Delta a$, $n_R$, $S_R$, and $L_R$. In the Methods section, we discuss how these parameters were chosen out of a large set (#) of possible descriptors. Compared to $n_R$, $S_R$, and $L_R$ which represent intrinsic properties of R, $\Delta a$, which is correlated with the crystallinity and the strain in the film, can be considered an extrinsic parameter.

We start by constructing a generalized linear model, elastic net (EN), which is a combination of Ridge and LASSO regressions[26]. This method assumes linear relationship between $S_{SSE}$ and the descriptors:

$$S_{SSE}(\Delta a, n_R, S_R, L_R) = \beta_0 + \beta_1 \Delta a + \beta_2 n_R + \beta_3 S_R + \beta_4 L_R \qquad (1)$$

Although linearity is a strong assumption, adopting it helps minimize over-fitting. Figure 3a shows the values of $\beta_1$, $\beta_2$, $\beta_3$ and $\beta_4$ obtained as a result of the regression fit. We can directly interpret the relationship between descriptors and $S_{SSE}$. The $n_R$ and $L_R$ have positive correlation with respect to $S_{SSE}$, while the $\Delta a$ and $S_R$ have negative correlation with $S_{SSE}$. However, EN regression analysis is limited to linear functions, and precludes proper modeling of more complicated dependencies. Therefore, to verify the patterns found by the EN, non-linear regression analysis using quadratic polynomial LASSO (QP-LASSO) model is applied next. We expand the linear model in equation 1 into a quadratic one:

$$\begin{aligned} S_{SSE}(\Delta a, n_R, S_R, L_R) = &\beta_0 + \beta_1 \Delta a + \beta_2 n_R + \beta_3 S_R + \beta_4 L_R \\ &+ \beta_5 \Delta a^2 + \beta_6 n_R^2 + \beta_7 S_R^2 + \beta_8 L_R^2 \\ &+ \beta_9 n_R S_R + \beta_{10} n_R L_R + \beta_{11} S_R L_R \end{aligned} \quad (2)$$

(Note that interaction terms between the extrinsic ($\Delta a$) and the intrinsic ($n_R$, $S_R$, $L_R$) factors with respect to the R element were not included.) QP-LASSO performs descriptor selection by adding $L_1$ regularization term. This term tends to suppress the coefficients ($\beta_0, ...., \beta_{11}$), and as a result only the ones in front of the most significant descriptors remain finite. Here, the QP-LASSO automatically selected four important descriptors: $\Delta a$, $n_R^2$, $S_R^2$ and $n_R L_R$. The values of these coefficients - $\beta_1$, $\beta_6$, $\beta_7$ and $\beta_{10}$ - are shown in Figure 3b. The $n_R$ and $n_R L_R$ terms are positively correlated with $S_{SSE}$, while the coefficients in front of $\Delta a$ and $S_R^2$ are negative. This agrees with the conclusions of the EN models.

The third algorithm we use is the NN - by far the most flexible of the three machine learning models we have employed. This comes at the price of significant risk of over-fitting, as well as difficulties interpreting the results. Figure 3c shows a visualization of the NN modeling result. It has input units (I, blue balls), hidden units (H, green balls), bias units (B, green balls) and an output unit (O, red balls). The strength of the dependence between units is represented as line width, with larger width indicating stronger connection. The red and blue lines denote positive and negative correlation between units, respectively. The relationship between input and output units is expressed as the product of the input-hidden correlation and hidden-output correlation. For example, I1-H4-O1 correlation is positive, because both I1-H4 (blue) and H4-O1 (blue) correlations are negative.

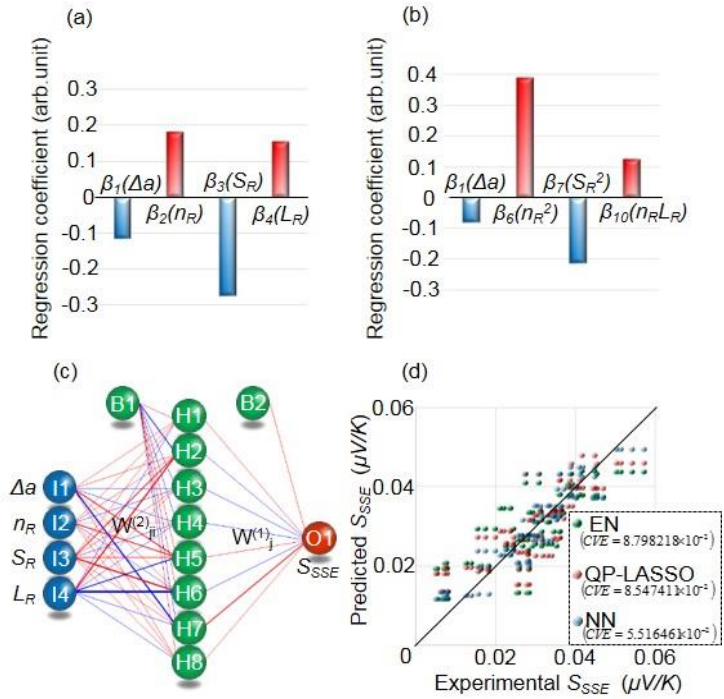

**Figure 3**

The visualization in Figure 3c provides a graphical summary of the relationship between the descriptors (the input units) and $S_{SSE}$ (the output unit). For $\Delta a$ (I1), the strongest path from I1 to O1 is I1-H7-O1 path, where I1-H7 connection is negative while H7-O1 connection is positive. This implies a negative correlation between $\Delta a$ and $S_{SSE}$. In case of $n_R$ (I2), the strongest path is I2-H5-O1 and both I2-H5 and H5-O1 connections are positive. Therefore, the relation between $n_R$ (I2) and $S_{SSE}$ is positive. For $S_R$ (I3), the strongest path from I3 to O1 is I3-H6-O1. This indicates overall negative correlation between $S_R$ (I3) and $S_{SSE}$ because I3-H6 connection is positive while H6-O1 is negative. Similarly, the relation between $L_R$ (I4) and $S_{SSE}$ is positive because of the strongest I4-H6-O2 path of which both I4-H6 and H6-O1 connections are negative.

Figure 3d shows the accuracy of EN, QP-LASSO, and NN models. The horizontal and vertical axes are the values of $S_{SSE}$ measured in the experiments and those predicted by the machine learning models, respectively.

We see that the NN model has better accuracy than the EN and QP-LASSO models, due to its much higher complexity. On the other hand, although the accuracy of the EN and the QP-LASSO models is not as high, interpreting their implications is much more straightforward. Despite these differences, the three machine learning algorithms converge on a picture where $S_{SSE}$ is positively correlated with $n_R$ and $L_R$, while negatively correlated with $\Delta a$ and $S_R$.

The positive correlation between $n_R$ and $S_{SSE}$ can be understood by considering the thermal conductivity $\kappa$ of the R:YIG film. $S_{SSE}$ is proportional to the temperature difference $\Delta T$ over the R:YIG layer [27], which in our setup is inversely proportional to $\kappa$. $\kappa$ is known to decrease with increasing $n_R$, due to the enhanced phonon scattering [28]. The negative correlation between $S_R$ and $S_{SSE}$ can be readily explained by considering $M_s$. The saturation magnetization decreases with increasing $S_R$ because the spin magnetic moment $S_R$ in R:YIG is arranged to cancel $M_s$ [29]. It is also known that the relation between $M_s$ and $S_{SSE}$ is positive: larger $M_s$ leads to larger spin currents [30]. Therefore, increasing $S_R$ lead to a drop in $S_{SSE}$. The negative correlation between $\Delta a$ and $S_{SSE}$ can be explained with the crystalline quality of R:YIG films. In general, larger lattice mismatch leads to worse crystallinity (through strain and defects), which in turn decreases the magnon (spin-current) diffusion length $D_L$ in the R:YIG film. In previous work it was found that $S_{SSE}$ decreases with decreasing $D_L$ [31]. It is thus straightforward to see how a larger $\Delta a$ (which results in local crystalline defects such as dislocations in the R:YIG layer) leads to a reduction in $S_{SSE}$.

Thus, the machine-learning derived correlations between $S_{SSE}$ and $\Delta a$, $n_R$, and $S_R$ can be explained based on the conventional understanding of the SSE physics. However, the positive correlation between $L_R$ and $S_{SSE}$ uncovered by the machine learning models appears to be beyond our current knowledge of SSE. The fact that increasing $L_R$ leads to increasing heat-to-spin current conversion efficiency cannot be easy reconciled with the magnon-driven theory of SSE [18], which has gained broad acceptance recently. In fact, within this theory, the opposite trend is expected; larger $L_R$ was believed to result in increase in the spin-phonon interaction, and accordingly reduction in magnon diffusion length[31] and thus smaller $S_{SSE}$. The positive correlation between $L_R$ and $S_{SSE}$ is more consistent with the phonon-driven SSE theory[19,32,33], where the phonon-mediated SSE may be enhanced by the large spin-phonon interaction due to $L_R$. In other words, increasing $L_R$ might increase the phonon contribution to the spin current generation. The surprising connection between $S_{SSE}$ and $L_R$, discovered by the machine learning models here, can perhaps lead to a more comprehensive description of the mechanism of SSE in the future.

**Development of superior STE materials through high-throughput experiments.** Although the exact origin of the relation between $L_R$ and $S_{SSE}$ cannot be unambiguously established at this time, we can nevertheless develop improved STE materials guided by this result. We apply this machine-learning-informed knowledge to develop anomalous Nernst effect (ANE) materials, in which part of the heat current could be converted into an electric current via thermally-driven electronic spin current. Since the ANE originates from the spin-orbit interaction, not only the SSE but also the ANE can be enhanced by tuning the orbital angular momentum $L_R$. The use of ferromagnetic alloy systems is preferable, as it limits the negative effect of increasing $L_R$; the conversion of spin current into electric current in the ANE occurs in the entire body of the material and is not affected by the drop in the spin-current diffusion length. This is in contrast to the SSE devices, where the conversion takes place only in the vicinity of the interface between the magnetic (e.g. R:YIG) and the paramagnetic (e.g. Pt) films.

Based on these considerations, we devise a simple design rule for novel ANE materials: we look for compounds which combine elements whose intrinsic properties can enhance the ANE voltage. In addition to the requirements of high saturation magnetization $M_s$ and large spin-orbit interaction (i.e., large atomic number element), we include elements with large orbital angular momentum $L_R$ (which according to the machine learning models is expected to increase the spin current generated by the heat current). Table 1 lists elements with large $M_s$, $Z$, and $L_R$. This simple design rule immediately leads to a large number of potential systems to search for multi-element compounds with enhanced ANE performance.

| $M_s$ ($\mu_B$) | $Z$ (1) | $L_R$ (1) |
|---|---|---|
| Fe (2.2) | Bi (83) | Pr (5) |
| Co (1.7) | Pb (82) | Nd (6) |
| Ni (0.6) | Tl (81) | Sm (5) |
|  | Au (79) | Dy (5) |
|  | Pt (78) | Ho (6) |
|  | Ir (77) | Er (6) |
|  | Os (76) | Tm (5) |

**Table 1**

As an initial example, we have selected the ternary Fe-Pt-Sm system. Because of the trade-offs between $M_s$, $Z$ and $L_R$ among these elements, it is important to optimize the composition within the ternary; for instance, increasing $M_s$ by using high Fe concentration can result in reduced $Z$ and $L_R$ because of low concentrations of Pt and Sm. To this end, we have carried out a high-throughput experiment [34-37] using a thin-film composition spread on SiO$_2$/Si, mapping a large fraction of the Fe-Pt-Sm ternary on one library wafer (Figure 4a, 4b). The thin film sample was then separated (Figure 4c) in order to measure the ANE thermopower $S_{ANE} (\equiv (V_{ANE}/\Delta T)(L_z/L_y))$ in the same manner as the above-described $S_{SSE}$ measurement. We apply the temperature difference $\Delta T$ and a magnetic field $H$ are applied along the z and x direction, respectively, and measure the anomalous Nernst voltage ($V_{ANE}$) along the y direction.

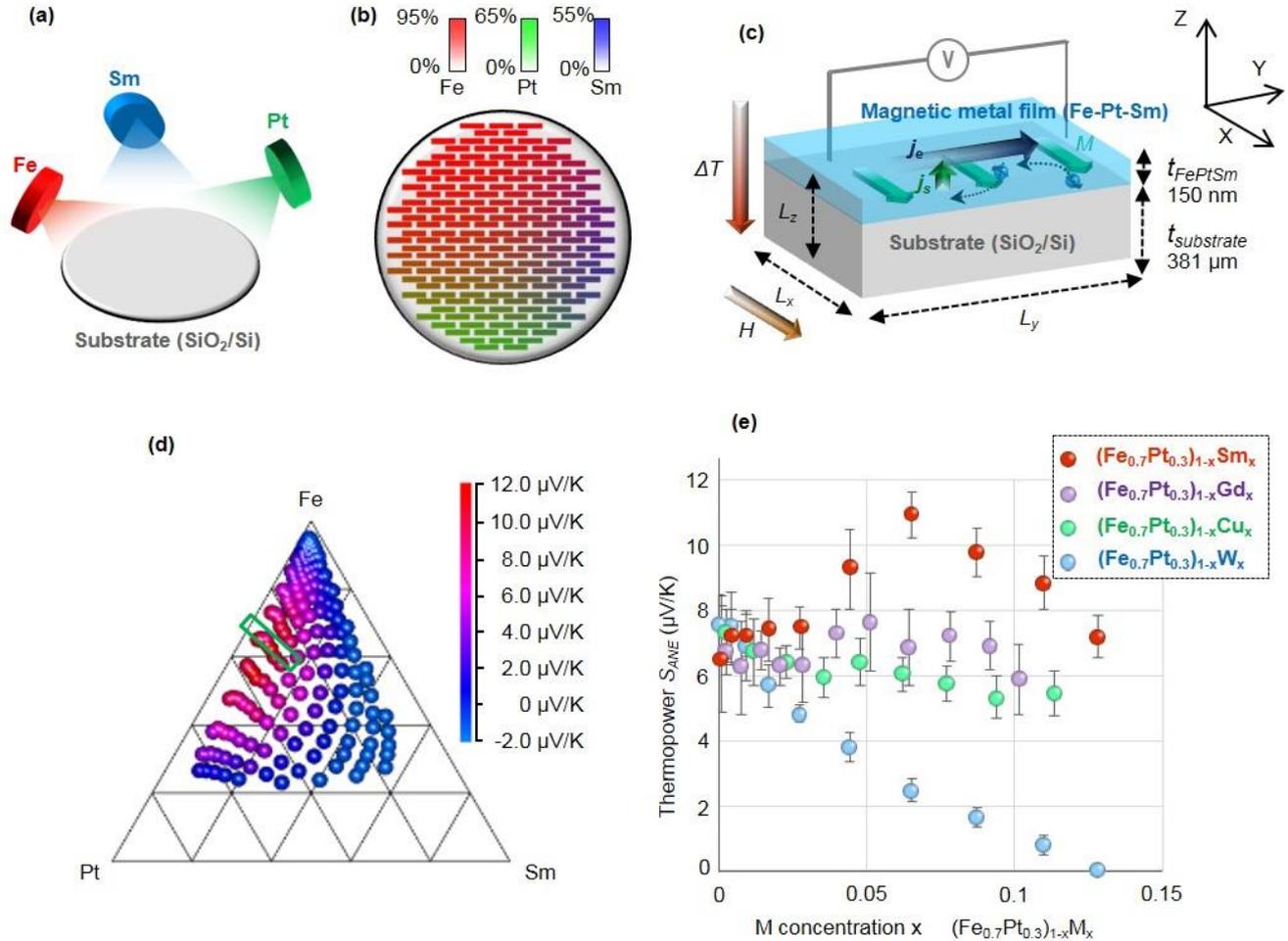

**Figure 4**

Figure 4d shows the measured $S_{ANE}$ values for the Fe-Pt-Sm spread film as a function of composition. The largest $S_{ANE}$ (peak) was detected near composition $Fe_{0.7}Pt_{0.3}Sm_{0.05}$. The existence of the peak agrees with the expected trade-off between $M_s$ (Fe), $Z$ (Pt) and $L_R$ (Sm). To further investigate the region outlined by the green rectangle in greater detail and confirm the contribution of $L_R$ (Sm), we fabricated $(Fe_{0.7}Pt_{0.3})_{1-x}M_x$ composition spread films on $SiO_2$/Si substrates by binary-combinatorial sputtering of $Fe_{0.7}Pt_{0.3}$ alloy with different metals (M =

Sm, Gd, Cu and W). Here, only Sm has finite orbital angular moment $L_R$, while other M elements have $L_R=0$.
Figure 4e shows the $V_{ANE}$ of the $(Fe_{0.7}Pt_{0.3})_{1-x}M_x/SiO_2/Si$. It is clear that a small amount of Sm is necessary to maximize $S_{ANE}$. When the Sm concentration x is less than about 6.5%, $S_{ANE}$ increases with Sm concentration due to its $L_R$ contribution. However, when Sm concentration exceeds 6.5%, the $S_{ANE}$ starts to decreases - the increase of $L_R$ associated with Sm cannot compensate the concomitant drop in $M_s$ and Z (due to reduction of Fe and Pt concentrations). With Gd, Cu and W, there is no $S_{ANE}$ enhancement since their $L_R$ is zero.

To benchmark the $S_{ANE}$ of the Fe-Pt-Sm material, we compare it with those of other ANE materials[38]. According to Ikhlas et al., $S_{ANE}$ of most ferromagnetic materials are on the order of 0.1 μV/K. In comparison, the largest $S_{ANE}$ obtained here (11.12 μV/K of $Fe_{0.665}Pt_{0.27}Sm_{0.065}$ in Figure 5e) is at least one order of magnitude larger than those of other known ANE materials. Note that the high-throughput investigation shows that the optimal Fe-Pt ratio occurs around $Fe_{0.7}Pt_{0.3}$ (with $S_{ANE}$ of ≈7 μV/K), which already gives a sizable improvement over the highest $S_{ANE}$ of the previously studied Fe and Pt materials (Pt/Fe multilayers and L10 FePt). This is then further significantly enhanced by the Sm substitution, as discussed above..

In summary, we have demonstrated the utility of machine learning both in exploring the fundamental physics of the STE phenomena and in optimizing the materials harnessing these effects. Combining it with high-throughput experimentation we have discovered an ANE material with $S_{ANE}$ an order of magnitude larger than that of any previously known ANE material. The data-driven approach has allowed us to construct unbiased statistical models for STE, which led us to a materials design rule, not rooted in the conventional theory of SSE. This design rule has, in turn, guided us to a family of new materials systems which exhibit large thermopower values. Further exploration of these ternary systems is underway and may provide the key to uncovering the fundamental mechanism of the STE phenomenon in the near future.

**Methods**

**Fabrication of spin-Seebeck devices (Pt/R:YIG/GGG or SGGG).** The fabrication method for the spin-Seebeck devices followed two steps. First, R:YIG layer was formed on the substrate (GGG or SGGG, 500 μm thickness) by means of the metal-organic-decomposition (MOD) method [39]. The MOD solution includes R, Y and Fe carboxylate, dissolved in organic solvents. Its chemical composition is R:Y:Fe = 1:2:5. The MOD solution was spin-coated on substrate at 1000 r.p.m. for 30 second, and then dried at 150 °C for 5 minutes. After pre-annealed at 450 °C for 5 minutes, it was annealed at 700 °C for 14 hours in air, to form a crystallized R:YIG layer. Its thickness was estimated to be 60 nm from the interference thickness meter. After completion of the R:YIG layer, a 10-nm-thickness Pt layer was deposited on the R:YIG layer by sputtering. For the measurement, the devices was cut into small chips, the length and width of which were 8.0 mm and 2.0 mm respectively.

**Fabrication of anomalous Nernst devices (Fe-Pt-Sm/SiO$_2$/Si).** The anomalous Nernst devices were fabricated as follows. The Fe-Pt-Sm film with composition gradient was deposited on 3 inch SiO$_2$/Si wafer by combinatorial sputtering at room temperature. The thickness of Fe-Pt-Sm, SiO$_2$ and Si layer are 150 nm, 0.5 μm and 381 μm, respectively. For the measurement, it was cut into small chips, with length and width identical to those cut from the Pt/R:YIG film.

**Measurement for the STE voltage ($V_{SSE}$ and $V_{ANE}$).** A temperature difference $\Delta T$ directed along the z direction as shown in figure 1a and figure 4c was applied between the top and the bottom of the devices, by sandwiching them between copper heat bath at 300 K and 300+$\Delta T$ K. The magnetic field H was applied along the x direction.

Under these conditions, the STE voltage ($V_{SSE}$ and $V_{ANE}$) can be detected along the y direction. The distance between voltage-detection terminals were set to 6 mm.

**Selecting descriptors and hyper-parameters for the machine learning models.** One major issue in developing machine learning models is avoiding overfitting. As a general rule, when the amount of available data is small, the number of descriptors should be constrained. For example, Seko et al.[40] employed machine learning to predict melting temperature, with the number of data points (compounds) and descriptors (predictors) 248 and 10, respectively. In our case there are only 112 data points (see Figure3d) and therefore it is advisable to use an even smaller number of descriptors. We have considered a number of descriptors, covering different properties of the rare-earth elements. These include atomic weight, spin and orbital angular momenta, number of unfilled orbitals, melting temperatures, magnetic moments, volumes and space groups (the last four are calculated for the elemental ground state). Magpie software was used to generate some of these [41]. Out of the more than 25 descriptors we have considered four were selected - $\Delta a$, $n_R$, $S_R$ and $L_R$. They are easy to interpret and connect to STE phenomenology, and at the same time models build with only these have accuracy comparable to that of models utilizing the full list of descriptors. In the future, we hope to increase the size of the experimental data and thus be able to include more descriptors in the model.

The hyper-parameters of the models are decided with the help of Leave-Out-One Cross-Validation (LOOCV) [26] – a widely used model optimization technique. In this scheme one data point is retained as validation data for testing the model, while the rest of the dataset is used as training data. The hyper parameters of the model are determined by minimizing the error indicator such as root mean square error (RMSE) or mean square error (MSE) on the test point. In this paper, we used RMSE as the error indicator. The LOOCV was carried out by "caret" package in R programming language.

**Elastic Net (EN)** [26]**.** The Elastic Net is a generalized linear model, combination of Ridge and Lasso regressions. The mixing ratio of the Ridge and the Lasso (Ridge : Lasso) was set to 1 : 0 based on the LOOCV. Therefore, in our case the EN model was equivalent to a Ridge regression. The LOOCV also decided the magnitude of generalization ($\lambda$ : 3.90625 × 10$^{-3}$), and the cross validation RMSE was 8.798218 × 10$^{-2}$. The EN was carried out by "glmnet" package in R programming language.

**Quadratic Polynomial Least Absolute Shrinkage and Selection Operator (QP-LASSO)** [26]**.** The LASSO is a regression analysis method that performs both variable selection and regularization. The QP-LASSO selects among quadratic, linear and constant terms. In this case QP-LASSO selected five valuables, including $\Delta a$, $n_R^2$, $S_R^2$ and $n_R L_R$, from equation (2). The LOOCV-determined magnitude of generalization is ($\lambda$ : 7.55559 × 10$^{-3}$), and the cross validation RMSE is 8.547411 × 10$^{-2}$. The QP-LASSO was carried out by "glmnet" package in R programming language.

**Neural Network (NN)** [26]**.** The NN method models the data by means of a statistical learning algorithm mimicking the brain. Here we have utilized simple 3-layer perceptron NN, with the number of input units, hidden units and output units being 4, 8 and 1, respectively. The hidden units and the output unit simulate the activation of a neuron by applying the hyperbolic tangent and the sigmoid functions, respectively. Mathematically, the NN models the non-linear function $S_{SSE}(\Delta a, n_R, S_R, L_R)$ by performing the following calculation.

$$S_{SSE}(\Delta a, n_R, S_R, L_R) = S_{SSE}(x, w) = \sigma\left(\sum_{j=0}^{8} w_j^{(2)} h\left(\sum_{i=0}^{4} w_{ij}^{(1)} x_i\right)\right)$$

*The $x_1$, $x_2$, $x_3$, $x_4$, $h$ and $\sigma$ are $\Delta a$, $n_R$, $S_R$, $L_R$, hyperbolic tangent function and sigmoid function, respectively. The* weights and the bias parameters $w^{(2)}_j$ and $w^{(1)}_{ji}$ are determined by minimizing the cost function with the backpropagation algorithm.[26] A decay value was set to $1.220703 \times 10^{-4}$. The hyper parameters, such as the number of hidden units and the value of the decay were decided by LOOCV, and cross validation RMSE of $5.516461 \times 10^{-2}$ was achieved. For NN analysis, we used "nnet" package in R programming language.

**Acknowledgements**

This work was supported by JST-PRESTO "Advanced Materials Informatics through Comprehensive Integration among Theoretical, Experimental, Computational and Data-Centric Sciences" (Grant No. JPMJPR17N4) and "Phase Interfaces for Highly Efficient Energy Utilization" (Grant No. JPMJPR12C1), CREST "Creation of Innovative Core Technologies for Nano-enabled Thermal Management" (Grant No. JPMJCR17I1), JST-ERATO "Spin Quantum Rectification Project" (Grant No. JPMJER1402), Grant-in-Aid for Scientific Research (A) (Grant No. JP15H02012), and Grant-in-Aid for Scientific Research on Innovative Area "Nano Spin Conversion Science" (Grant No. JP26103005) from JSPS KAKENHI, Japan. I.T. is supported in part by C-SPIN, one of six centers of STARnet, a Semiconductor Research Corporation program, sponsored by MARCO and DARPA.

**Author contributions**

Y.I., M.I., A.K., K.T., H.S. and K.I. designed the experiment, fabricated the samples, and collected all of the data. Y.I., R.S., K.U. and E.S. contribute to the theoretical discussion. Y.I., R.S., V.S., A.G.K. and I.T discussed the results of machine learning modeling. Y.I., V.S., I.T. and M.I wrote the manuscript. S.Y. supervised this study. All the authors discussed the results and commented on the manuscript.

**Additional information**

The authors declare no competing financial interests.

**Figure legends**

**Figure 1 | A spin-Seebeck device. a**, Schematic of the spin-Seebeck device consisting of a Pt layer, a rare-earth substituted yttrium iron garnet ($R_1Y_2Fe_5O_{12}$, referred to as R:YIG) layer and a (111)-oriented Gadolinium Gallium Garnet ($Gd_3Ga_5O_{12}$, referred to as GGG) substrate or a (111)-oriented Substituted Gadolinium Gallium Garnet ($Gd_{2.675}Ca_{0.325}Ga_{4.025}Mg_{0.325}Zr_{0.65}O_{12}$, referred to as SGGG) substrate. Spin current $j_s$ generated from heat current in the magnetic layer by the spin-Seebeck effect is converted into electrical current $j_e$ by the inverse spin Hall effect in the Pt layer. **b,** Cross-section transmission electron microscopy image of a spin-Seebeck multilayer structure (Pt/YIG/SGGG). The inset shows that the magnetic layer has grown epitaxially on the substrate. **c,** The spin-Seebeck voltage $V_{SSE}$ of a spin-Seebeck device (Pt/YIG/SGGG) as a function of $H$. The sign of $V_{SSE}$ follows the sign of $H$, indicating that the thermo-electromotive force arises from the spin-Seebeck effect and the inverse spin-Hall effect.

**Figure 2 | SSE thermopower $S_{SSE}$ for different rare-earth substituted YIG (R:YIG).** The spin-Seebeck thermopower $S_{SSE}$ for different rare-earth substituted YIG (R:YIG). The magnitude of $S_{SSE}$ varies depending on the choice of rare-earth element. Error bars are standard deviation.

**Figure 3 | Informatics approach. a.** Regression coefficients for the elastic net (EN) model. The value of constant term $β_0$ is 0.3310662. $\Delta a$ and $S_R$ are negatively correlated with $S_{SSE}$, while the $n_R$ and $L_R$ have positive correlation with $S_{SSE}$, **b.** Regression coefficient in the quadratic polynomial LASSO (QP-LASSO). The value of $β_0$ is 0.3039554. $\Delta a$ and $S_R^2$ are negatively correlated with $S_{SSE}$, while the $n_R^2$ and $n_RL_R$ have positive correlation with

$S_{SSE}$, **c.** Visual representation of the neural network (NN) model. The line width represents the connection strength between units. Red/blue color demonstrate positive/negative correlation. **d.** Predicted vs. measured values of $S_{SSE}$ for the EN, QP-LASSO and NN models. The value of cross validation error (CVE) are shown in the legend.

**Figure 4 | Development of better STE (ANE) materials. a.** Schematic illustration of combinatorial sputtering method. **b.** Composition spread mapping of Fe-Pt-Sm film fabricated by combinatorial sputtering. **c.** Schematic of the anomalous Nernst device consisting of a Fe-Pt-Sm layer and $SiO_2$/Si substrate. **d.** Anomalous Nernst thermopower $S_{ANE}$ of Fe-Pt-Sm/$SiO_2$/Si as a function of composition data. **e.** Anomalous Nernst voltage $S_{ANE}$ of $(Fe_{0.7}Pt_{0.3})_{1-x}M_x$/$SiO_2$/Si as a function of composition data. Error bars show standard deviations.

**Table 1 | Lists of elements with large $M_s$, $Z$, and $L_R$ for designing new spin-driven thermoelectric materials.** We list Fe, Co, and Ni as examples of high $M_s$ ferromagnetic elements (the number in the parenthesis is magnetic moment in $\mu_B$, the Bohr magneton units). The elements with large atomic number $Z$ are listed. For high orbital angular momentum $L_R$ elements, ones with dimensionless quantum number $L_R$ larger than or equal to 5 are listed. We propose a simple design rule where we construct ternary compounds A-B-C, where A, B, and C are elements with large $M_s$, $Z$, and $L_R$, respectively. Our initial ternary example is Fe-Pt-Sm.